\documentclass[11pt,a4paper]{article}
\usepackage[T1]{fontenc}
\usepackage[margin=1in]{geometry}
\usepackage{lmodern,microtype}
\usepackage{authblk}
\usepackage{amsmath,amssymb,bm}
\usepackage{graphicx,xcolor,caption,booktabs,tabularx}
\usepackage{float}
\usepackage[section]{placeins}
\usepackage[
    backend=biber,
    style=phys,
    sorting=none,
    biblabel=brackets,
    articletitle=true,
    pageranges=false,
    maxbibnames=10,
    minbibnames=10,
    eprint=true,
    doi=false,
    url=false
  ]{biblatex}

\DeclareFieldFormat*{title}{#1}

\DeclareFieldFormat[misc,unpublished]{date}{%
\iffieldundef{eprint}{#1}{\mkbibparens{#1}}%
}

\AtEveryBibitem{%
    \iffieldundef{journaltitle}{%
      \iffieldundef{booktitle}{}{%
        \clearfield{eprint}%
        \clearfield{eprinttype}%
        \clearfield{eprintclass}%
      }%
    }{%
      \iffieldequalstr{journaltitle}{arXiv}{}{%
        \clearfield{eprint}%
        \clearfield{eprinttype}%
        \clearfield{eprintclass}%
      }%
    }%

    \iffieldundef{eprint}{}{%
      \clearfield{month}%
      \clearfield{day}%
      \clearfield{eprintclass}%
    }%
}
\usepackage[colorlinks=true,linkcolor=blue!45!black,citecolor=blue!45!black,urlcolor=blue!45!black]{hyperref}
\newcommand{\CT}{\mathcal{T}}

\providecommand{\doi}[1]{\href{https://doi.org/#1}{doi:\nolinkurl{#1}}}

\title{CT-bMPS: Approximate Tensor Network Contraction with Boundary MPS on Contraction Trees}
\author[1]{Hidetaka Manabe\thanks{Email: \href{mailto:hidetaka_manabe@sutd.edu.sg}{\nolinkurl{hidetaka_manabe@sutd.edu.sg}}}}
\author[1]{Feng Pan}
\affil[1]{Science, Mathematics and Technology Cluster, Singapore University of Technology and Design (SUTD), Singapore}
\date{}
\hypersetup{pdftitle={CT-bMPS: Approximate Tensor Network Contraction with Boundary MPS on Contraction Trees},pdfauthor={Hidetaka Manabe and Feng Pan}}

\begin{document}
\maketitle
\begin{abstract}
We introduce contraction tree boundary matrix product states (CT-bMPS), a scalable approximate contraction method for general closed tensor networks. CT-bMPS represents the boundary tensor and its complementary environment at each edge of a contraction tree as matrix product states. This representation allows local compression based on a reduced transition matrix to incorporate environment information along arbitrary contraction trees, while keeping the time and memory costs polynomial in the network size and bond dimension. We benchmark the method on a variety of models and applications, including Ising partition functions, random tensor networks, kicked Ising dynamics, and quantum error correction decoding. These benchmarks demonstrate improved accuracy and computational efficiency over existing general-purpose approximate contraction methods, together with the ability to handle larger and more complex networks. Optimized contraction trees and environment updates substantially improve accuracy without increasing the bond dimension. These results establish CT-bMPS as a general and scalable approximation framework with flexible contraction orders and environment-based compression for a wide range of applications.

\end{abstract}

\section{Introduction}
\label{sec:introduction}

Tensors, or multi-index arrays, are fundamental objects in physics and computer science. Explicitly storing these objects can require exponentially many parameters, a difficulty known as the curse of dimensionality. Tensor networks (TN) address this challenge by expressing a large tensor as a network of smaller tensors connected by indices~\cite{orus2014,oseledets2011}. When the tensor admits such a representation with modest bond dimensions, a TN can avoid the exponential storage cost of the full array. TN representations arise naturally when the state space has a tensor product structure and interactions act on only a few subsystems. Tensor networks have been used in a broad range of fields, including statistical mechanics~\cite{xie2012}, quantum many body physics~\cite{vidal2004,verstraete2004}, quantum circuit simulation~\cite{markovshi2008}, high-energy physics~\cite{swingle2012,pastawski2015}, and large-scale data analysis~\cite{cichocki2016,stoudenmire2016,han2018}.

However, representing a tensor as a TN does not by itself make its contraction computationally tractable. Contracting a TN exactly, by summing over all internal indices, is \#P-hard in general~\cite{schuch2007}. Two approaches have been developed to address this difficulty. The first approach is to reduce the cost of exact contraction by optimizing the \emph{contraction tree}. A contraction tree is a binary tree that specifies the order of pairwise tensor contractions. The choice of contraction tree can drastically change both the time and space complexity. Although finding a tree that minimizes the cost is itself NP-hard~\cite{bridgeman2017}, efficient heuristic algorithms have been developed to find low-cost trees for a given tensor network~\cite{huang2021,panzhang2021,graykourtis2021}. Moreover, we can combine these optimized contraction trees with parallel computation on supercomputers and GPUs to accelerate tensor network contraction~\cite{villalonga2019,huang2021,pangpu2024}. This approach is one of the leading methods for classically simulating random circuit sampling~\cite{panzhang2021,panchen2022} and is widely used to establish classical baselines for quantum computational advantage experiments~\cite{arute2019,morvan2024}.

\begin{figure}[t]
\centering
\includegraphics[width=.94\linewidth]{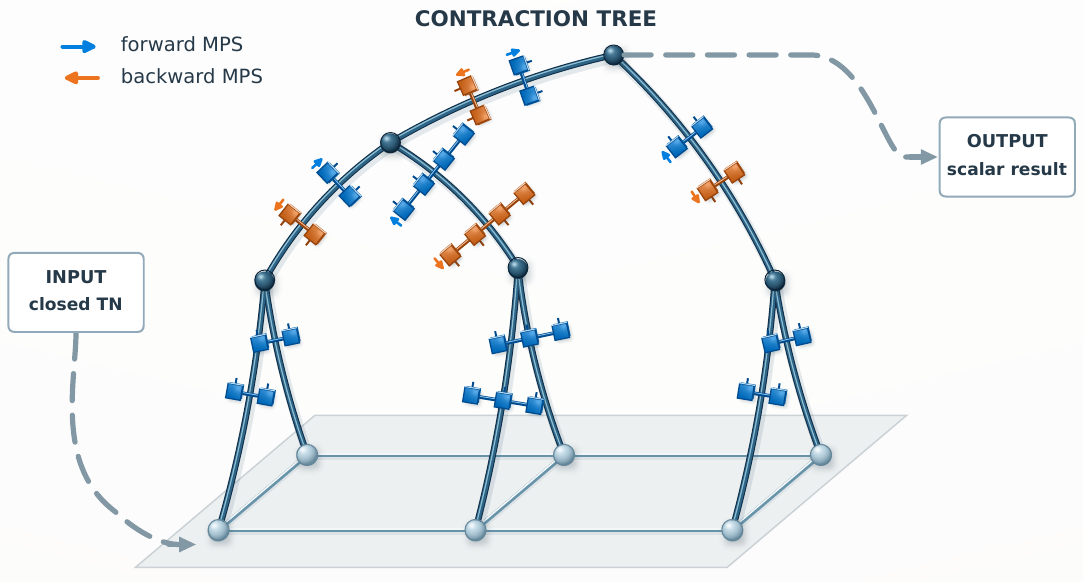}
\caption{CT-bMPS overview. The input TN is shown on the lower plane; the contraction tree above it groups subsets of its tensors. Each tree edge carries a boundary MPS rather than a dense tensor. Messages in both directions represent the two complementary parts of the TN. The opposite message serves as an environment for local truncation, and pairing the two messages gives the scalar.}
\label{fig:overview}
\end{figure}

The second approach is to contract the TN approximately. When exact contraction exceeds the available computational resources, approximation becomes necessary to evaluate larger TNs. Tensor network ansatzes, such as matrix product states (MPS) and projected entangled pair states (PEPS), have long been used to approximate quantum many-body states~\cite{schollwoeck2011,cirac2021}. In boundary MPS methods, an MPS is used to approximate the tensor on the boundary between the contracted and uncontracted parts of a TN~\cite{verstraete2004,orus2014,chubb2021}. The MPS is updated by successively absorbing tensors as the contraction proceeds. Its bond dimensions can be controlled by using the canonical form and singular value decomposition (SVD) to discard small singular values~\cite{schollwoeck2011}. Incorporating information from the remaining tensors, called the environment, can further improve accuracy by accounting for how the retained components contribute to the final contraction~\cite{xie2009,evenbly2018}.

However, these TN approximations are often designed separately for particular problems, geometries, or ansatzes. For example, the tensor renormalization group contracts regular lattice TNs through coarse-graining steps adapted to the lattice geometry~\cite{levinnave2007}. In quantum many-body physics, the choice of TN ansatz is commonly guided by the entanglement structure of the quantum states~\cite{cirac2021}. In data analysis, TN approximations can exploit weak correlations between variables in classical data~\cite{oseledets2011,manabe2025}. As finite TNs are used across increasingly diverse fields, it becomes important to approximate them without redesigning the approximate scheme for each problem. Several general methods pursue this direction~\cite{pan2020,gray2024,ma2024}. A central challenge is to develop a scalable approximation framework for general TNs that supports flexible contraction orders and environment-based compression.

Here we propose \emph{contraction tree boundary MPS} (CT-bMPS), an approximate contraction method for general closed TNs, including TNs with hyperedges. CT-bMPS combines contraction trees used in exact contraction, boundary MPS representations of intermediate tensors, and compression based on a reduced transition matrix (RTM). Figure~\ref{fig:overview} gives an overview. We first construct a contraction tree for the input TN and represent the boundary tensor at each tree edge by an MPS. Starting from the leaves, two incoming boundary MPS are combined at each internal node, with compression to a maximum bond dimension $\chi$ during the contraction. Repeating this operation along the tree gives an approximation to the full contraction. We refer to this procedure as \emph{ordinary CT-bMPS}.

The same contraction tree also allows information to travel in the opposite direction. A backward MPS represents the complementary part of the TN and supplies an environment for RTM-based compression when combining two forward MPS~\cite{hastings2015,li2026,ouyang2026}. This environment guides the local truncation by accounting for how the retained components contribute to the full contraction. Sweeping through the tree in both directions updates the boundary MPS messages iteratively. We refer to this procedure as \emph{environment CT-bMPS}.

Our method runs in polynomial time and memory in the TN size and the maximum MPS bond dimension $\chi$, enabling scalable approximate contraction of general TNs. Its accuracy depends strongly on the contraction tree and the index order within each boundary MPS. These choices determine which intermediate tensors are approximated and how their indices are divided at each MPS cut. We use simulated annealing (SA) to search for trees with low total exact-contraction FLOPs~\cite{morvan2024} and construct a global index order from the geometry of the TN. For our benchmark models, the SA-optimized trees can improve both approximation accuracy and computational cost over geometric-sweep orders.

We evaluate the accuracy, scalability, and generality of CT-bMPS by applying the same approximate contraction scheme to Ising partition functions, random TNs, kicked Ising dynamics~\cite{kim2023}, and quantum error correction decoding~\cite{piveteau2024}. These benchmarks demonstrate that CT-bMPS combines high accuracy with improved scalability to larger systems compared with existing approximate contraction methods. In particular, environment updates substantially improve approximation accuracy at the same bond dimension. In kicked Ising simulations, CT-bMPS yields estimates close to classical reference values at much smaller bond dimensions, demonstrating its applicability to quantum circuit simulation. Finally, we apply CT-bMPS to approximate maximum-likelihood decoding of surface and color codes up to distance $d=7$. Our decoding comparisons show lower logical error rates than the reference decoders, supporting the use of CT-bMPS to investigate the error-correction performance of QEC codes. These results demonstrate that, without redesigning the approximate scheme for each problem, flexible contraction orders and environment-based compression support accurate and scalable approximation across different applications.

\subsection{Related work}
Several methods have been proposed for the approximate contraction of general TNs~\cite{pan2020,gray2024,ma2024}. Pan \emph{et al.} introduced CATN for approximate contraction of TNs with arbitrary connectivity~\cite{pan2020}. They represent input and intermediate tensors as MPS and combine index swaps, contractions, and index merges according to a chosen contraction order. CT-bMPS can be viewed as a natural extension of CATN. We organize the MPS as boundary messages on an explicit contraction tree and handle hyperedges directly. Each boundary MPS can also be updated using the opposite message as its environment.

Gray and Chan's hyperoptimized approximate contraction (HAC) also optimizes contraction trees and truncates bond dimensions during contraction to limit the growth of intermediate tensors~\cite{gray2024}. They use a local spanning tree of the current TN as an approximate environment to improve truncation accuracy. Whereas HAC represents intermediate tensors as dense tensors, CT-bMPS represents them as MPS. At fixed MPS bond dimension, the cost of storing and approximately contracting these tensors grows polynomially rather than exponentially with the number of boundary indices.

Ma \emph{et al.} formulate a flexible approach to approximate contraction using a contraction tree over a partitioned TN~\cite{ma2024}. At each step, they embed the union of two partitions into a chosen tree structure and use density-matrix compression to obtain a TTN with small bond dimensions, such as an MPS or comb. Their analysis does not establish a general polynomial-cost bound in the TN size and output bond dimension for this embedding-based compression. CT-bMPS instead specifies local MPS contraction and truncation steps along the contraction tree, giving polynomial time and memory in the TN size and MPS bond dimension. The environment also differs: Ma's compression uses information within the two partitions being combined, whereas CT-bMPS uses the opposite MPS message to incorporate the complementary part of the full TN.

Temporal contraction and folding methods also motivate our use of environments~\cite{banuls2009,hastings2015}. In particular, Ouyang \emph{et al.} use an opposite temporal MPS when compressing the product of a temporal MPS and an MPO, and update the environments through sweeps~\cite{ouyang2026}. Our environment CT-bMPS applies this idea to boundary MPS compression at each internal node of an arbitrary contraction tree. Repeating these updates in both directions extends this truncation procedure from a temporal chain to a general contraction tree.

\subsection{Structure of the paper}
This paper is organized as follows. Section~\ref{sec:method} describes the CT-bMPS method. Section~\ref{sec:benchmark-ising} benchmarks CT-bMPS on Ising partition functions and random TNs. Section~\ref{sec:benchmark-ibm} studies the classical simulation of quantum circuits implementing kicked Ising dynamics. Section~\ref{sec:benchmark-decoding} applies the method to decoding quantum low-density parity-check codes, focusing on surface and color codes. Section~\ref{sec:discussion} discusses the advantages and limitations of CT-bMPS and directions for future work.

\section{Method}
\label{sec:method}
In this section, we describe the CT-bMPS algorithm, a method for the approximate contraction of general closed TNs. We first define the TN and its contraction tree, then explain the ordinary contraction, the environment update, and the selection of the contraction tree and index order.

\subsection{Closed tensor networks and contraction trees}
\label{sec:tn-and-trees}
Let $V$ be a finite set of tensors and $\mathcal I$ a set of indices. Index $i$ takes values in a domain $D_i$ of size $d_i$, and tensor $A_v$ has index set $I(v)\subseteq\mathcal I$. A closed TN represents the scalar
\begin{equation}
 Z=\sum_{\{x_i\in D_i:i\in\mathcal I\}}\prod_{v\in V}A_v(x_{I(v)}).
 \label{eq:closed}
\end{equation}
Indices shared by more than two tensors are called hyperedges.

A binary contraction tree $\CT$ has the input tensors as leaves and degree-three internal vertices. The tree encodes the order of pairwise tensor contractions, and each internal vertex corresponds to the intermediate tensor produced by one such contraction~\cite{graykourtis2021}. Choosing an edge as the root specifies the final pairwise contraction. The remaining contractions proceed from the leaves toward this root.

Removing an edge of $\CT$ separates the input tensors into two sets $S$ and $\bar S=V\setminus S$. Their boundary $B(S)$ consists of indices occurring on both sides. Contracting the tensors in $S$, and summing only indices whose occurrences all lie in $S$, gives a boundary tensor $m_S$. We call this tensor a message directed toward the other side of the edge. The opposite message $m_{\bar S}$ has the same boundary indices, and their pairing recovers the full contraction,
\begin{equation}
 Z=\sum_{x_{B(S)}}m_S(x_{B(S)})m_{\bar S}(x_{B(S)}).
 \label{eq:root-pairing}
\end{equation}
\subsection{Ordinary CT-bMPS}
\label{sec:ordinary}
\label{sec:index-order}
Ordinary CT-bMPS follows the contraction tree while representing each boundary tensor as an MPS. For an ordered boundary $(i_1,\ldots,i_p)$, we write
\begin{equation}
 \widehat m_S(x_{i_1},\ldots,x_{i_p})
 =\sum_{\alpha_1,\ldots,\alpha_{p-1}}
 C^{[1]x_{i_1}}_{\alpha_1}
 C^{[2]x_{i_2}}_{\alpha_1\alpha_2}
 \cdots C^{[p]x_{i_p}}_{\alpha_{p-1}} .
 \label{eq:mps}
\end{equation}
Here $C^{[k]}$ is the tensor at site $k$, and $\alpha_k$ are virtual indices. The virtual bond dimensions are bounded above by $\chi$. We fix a global index order, defined as a total order on the full index set $\mathcal I$. The boundary order at each edge of the contraction tree is obtained by restricting this order to its boundary indices, ensuring consistent index ordering across all edges. The choice of the global index order is discussed in Sec.~\ref{sec:tree-selection}.

\begin{figure}[t]
\centering
\includegraphics[width=\linewidth]{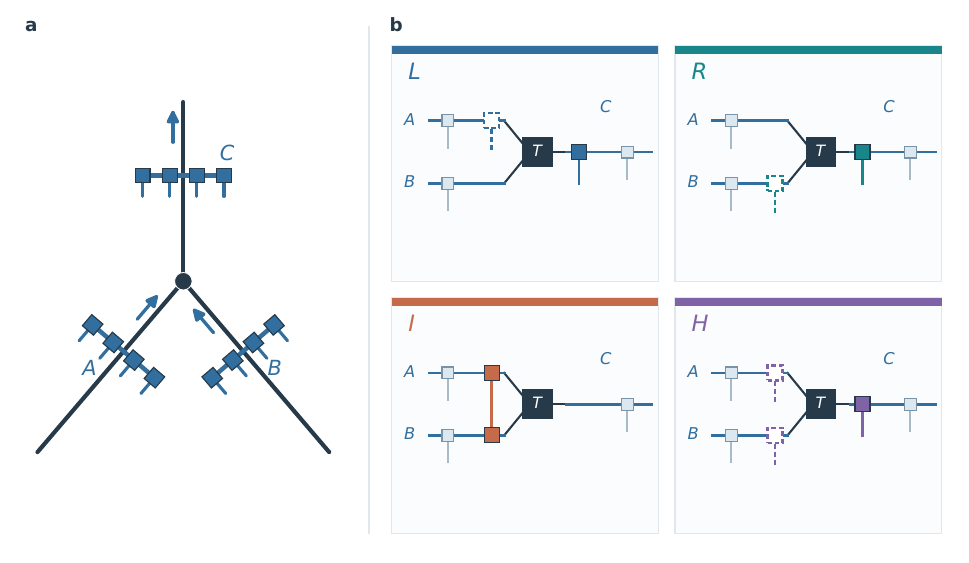}
\caption{Ordinary CT-bMPS. (a) Two incoming boundary MPS messages $A$ and $B$ form an outgoing message $C$ at a contraction tree node. (b) Four local index patterns around the core tensor $T$. Indices $L$ and $R$ occur in only one input and survive in $C$. An internal index $I$ is summed and has no output site. A shared hyperedge $H$ with an outside owner survives as one site of $C$. Dashed colored input sites indicate transfer into the solid output site.}
\label{fig:ordinary-method}
\end{figure}

Starting from the leaf tensors, we repeatedly combine two incoming MPS into one outgoing MPS. Figure~\ref{fig:ordinary-method}(a) illustrates that two incoming MPSs, $A$ and $B$, are contracted into an outgoing MPS $C$ at an internal node of the contraction tree. Panel (b) shows the four possible index patterns that arise when contracting the two MPS. $L$ denotes indices that occur only in the first input and remain in the output. $R$ denotes indices that occur only in the second input and also remain in the output. $I$ denotes indices shared by both inputs but absent from the outgoing MPS; these indices are summed over. $H$ denotes indices shared by both inputs and retained in the outgoing MPS. The local contraction is represented as
\begin{equation}
 C(x_L,x_H,x_R)=\sum_{x_I}A(x_L,x_H,x_I)B(x_I,x_H,x_R).
 \label{eq:local-product}
\end{equation}
In particular, a hyperedge is not eliminated merely because it occurs in both inputs.

We first put the input MPS $A$ and $B$ in left-canonical form without truncation~\cite{schollwoeck2011}. We then use a core tensor $T$ to transform $A$ and $B$ into the output MPS $C$ one site at a time, following the global index order from right to left. The role of $T$ depends on the four index patterns in Fig.~\ref{fig:ordinary-method}(b). For an index in $L$, we absorb the corresponding site tensor of $A$ into $T$, then use an SVD to split off a site tensor of $C$, leaving an updated core tensor $T$. For an index in $R$, we perform the same operation using the corresponding site tensor of $B$. For an index in $I$, both inputs have a corresponding site tensor. We absorb both into $T$ and sum over their shared index, without creating an output site. For an index in $H$, we absorb both site tensors into $T$ by taking their outer product at each fixed value of the shared index. This index is retained once rather than summed over, and an SVD again splits off a site tensor of $C$.

With input and output virtual bond dimensions bounded by $\chi$ and maximum physical dimension $d$, each SVD acts on a matrix of size at most $\chi^2\times d\chi$ and costs $O(d^2\chi^4)$ time. For binary indices, this is a $\chi^2\times2\chi$ matrix with an $O(\chi^4)$ SVD cost. At fixed physical dimension, this procedure approximately contracts two MPS sharing a global index order in $O(h\chi^4)$ time, where $h$ is the number of indices processed.

\subsection{Environment CT-bMPS}
\label{sec:environment}
Every edge of an unrooted contraction tree separates two complementary parts of the TN. Alongside a forward boundary MPS, we can therefore construct a backward MPS representing the other part. This opposite directed message provides an approximation to the full environment of the forward message. We use it to improve local truncation through the RTM construction~\cite{hastings2015,ouyang2026}.

\begin{figure}[t]
\centering
\includegraphics[width=\linewidth]{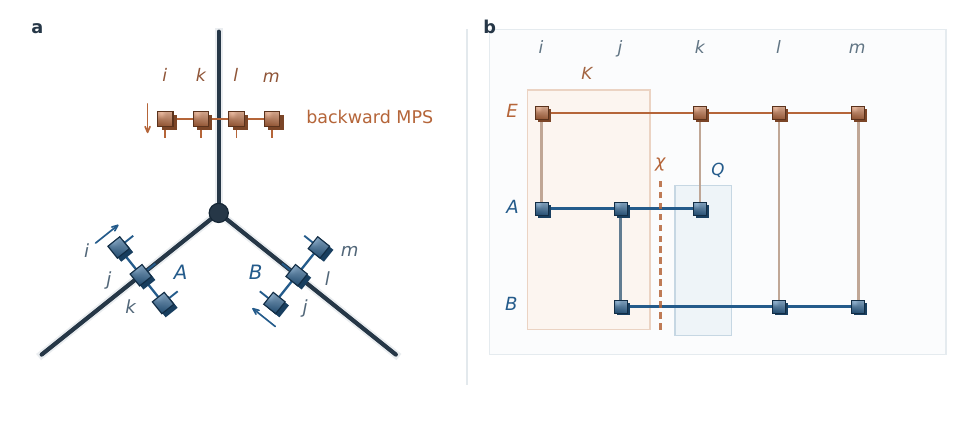}
\caption{Environment CT-bMPS. (a) The opposite boundary MPS provides the environment $E$ for combining the incoming messages $A$ and $B$. (b) Local truncation at the cut between $j$ and $k$. Contracting the tensors to the left of the cut gives $K$, while $Q$ is formed from the core tensor $T$.}
\label{fig:environment-method}
\end{figure}

We initialize the directed MPS messages by ordinary contraction, with an initial bond cap $\chi_0$. To construct the output MPS $C$ using the environment, we put the input MPS $A$ and $B$ and the environment MPS $E$ in canonical form. We keep $E$ fixed during this contraction.

As in ordinary CT-bMPS, we construct $C$ from right to left using the core tensor $T$. At the current site, we absorb the corresponding tensors of $A$ and $B$ into $T$. For an index in $L$, $R$, or $H$, we reshape $T$ into a matrix $Q$ and use the environment to guide its truncation. To form $K$, we contract the unprocessed tensors of $A$ and $B$ with the corresponding tensors of $E$. The virtual bonds at the current cut remain open. We then apply an SVD to $KQ$ to determine the next site tensor of $C$ and update $T$. Figure~\ref{fig:environment-method} shows an example.

We update the MPS messages by sweeping through the entire contraction tree. These sweeps provide bidirectional MPS messages, whose contraction at a tree edge gives an estimate of the scalar $Z$ through Eq.~\eqref{eq:root-pairing}. We use the average of the edge estimates collected during the final sweep unless a selected-root estimate is specified for the application. Increasing the number of sweeps allows us to examine whether the reported estimate converges. This convergence test is only a heuristic indicator: neither convergence of the estimate nor convergence to the exact value is guaranteed.

For $N$ input tensors of bounded degree, maximum physical dimension $d$, and MPS bond dimensions bounded by $\chi$, each local RTM update takes $O(d\chi^4)$ time, with the leading cost coming from matrix multiplications that form $KQ$ and update the core tensor. Each edge update processes at most $O(N)$ indices, and a sweep visits $O(N)$ tree edges, giving $O(N^2d\chi^4)$ time for a sweep of RTM updates. Storing the bidirectional MPS messages requires $O(N^2d\chi^2)$ space. Both time and space therefore scale polynomially. The RTM SVD itself acts on a matrix of size at most $\chi\times d\chi$ and costs $O(d\chi^3)$ time. Inactive messages and cached tensors could be kept in host memory or on disk and loaded when needed, reducing device-memory requirements and facilitating a GPU implementation.

\subsection{Contraction tree and index-order optimization}
\label{sec:tree-selection}
The accuracy of CT-bMPS depends strongly on the contraction tree. Optimizing contraction trees is a well-developed approach to reducing the cost of exact contraction~\cite{huang2021}. Following the use of simulated annealing (SA) for contraction-order search in Ref.~\cite{morvan2024}, we use SA to search for trees with a low exact-contraction FLOP count. We use this objective because the sizes of intermediate tensors and the number of local truncation steps affect approximation accuracy. Smaller intermediate tensors can be easier to approximate at a fixed bond dimension, while fewer truncations can reduce the accumulation of approximation errors. However, this objective does not directly account for approximation error. Developing contraction tree optimization specifically for CT-bMPS remains future work.

After selecting a tree, we fix its root and construct a global index order from the structure of the input TN. Every boundary, including the opposite environment message, uses a restriction of this same order. The global index order strongly affects performance because the accuracy of a boundary MPS at a fixed bond dimension depends on its index ordering. In this work, we define the global index order by sorting the indices according to their geometric coordinates, comparing the first coordinate, then the second, and so on. This choice aims to place geometrically nearby indices close together along the boundary MPS and is expected to improve approximation accuracy.

\section{Ising partition functions and random tensor networks}
\label{sec:benchmark-ising}
To evaluate the performance of CT-bMPS, we study the Ising model and random TNs and compare our method with existing approximate contraction methods. All calculations in this and the following sections use double precision, one CPU worker, and one BLAS thread per run.

\subsection{Cubic Ising model and benchmark setup}
We consider the ferromagnetic Ising model on a cubic lattice of side \(L\) with open boundary conditions,
\begin{equation}
 Z=\sum_{\{\sigma_u=\pm1\}}
 \prod_{\langle uv\rangle} \exp(\beta\sigma_u\sigma_v),
 \label{eq:ising-z}
\end{equation}
with \(\beta=0.3\). Each interaction is represented by a two-index tensor and each spin by a shared binary index. An interior spin connects six tensors, giving a native hyperedge representation without auxiliary copy tensors.

\begin{figure}[t]
\centering
\includegraphics[width=0.80\linewidth]{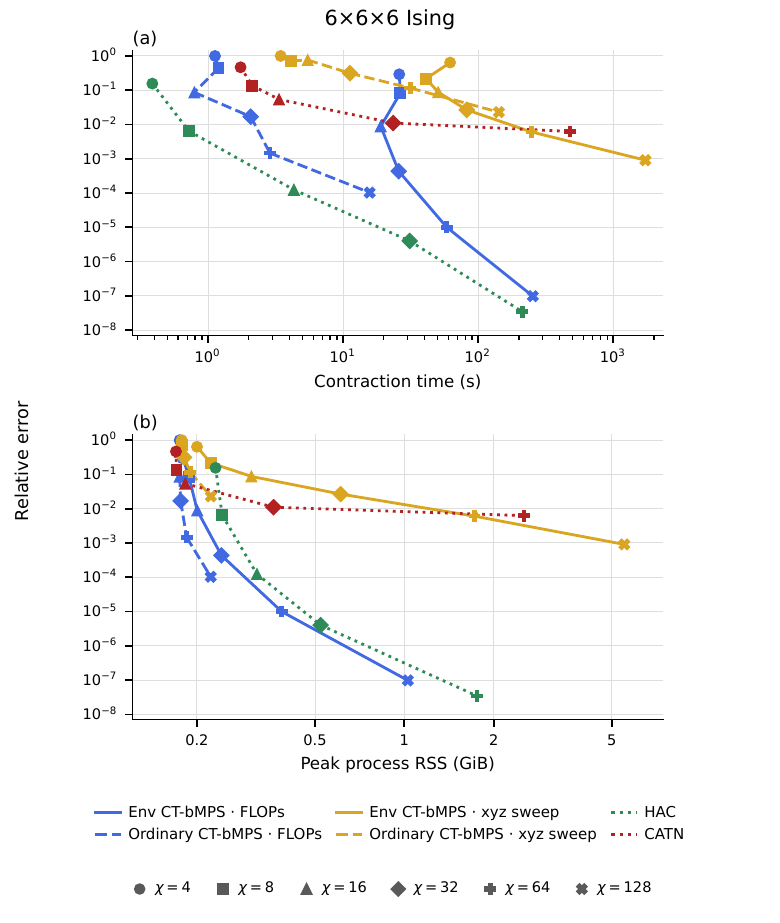}
\caption{The \(6\times6\times6\) Ising model at \(\beta=0.3\). Relative scalar error is plotted against (a) contraction time and (b) peak process RSS. FLOPs denotes the CT-bMPS path obtained by simulated annealing to minimize exact contraction FLOPs; \(xyz\) denotes a geometric sweep of the lattice. Markers indicate \(\chi\). CT-bMPS uses native spin hyperedges; HAC and CATN use pairwise representations.}
\label{fig:ising}
\end{figure}

We compare a contraction tree optimized by SA for the FLOP count of exact contraction with a geometric \(xyz\)-sweep tree. Both use a common global index order based on the lattice geometry. Ordinary CT-bMPS uses the balanced root to estimate $Z$. Environment CT-bMPS initializes the message bank at the same \(\chi\), performs five sweeps, and uses the average estimate described in Sec.~\ref{sec:environment}.

The external comparisons are CATN~\cite{pan2020} and hyperoptimized approximate contraction (HAC)~\cite{gray2024}. Both use pairwise vertex-tensor representations of the same physical Ising model, whereas CT-bMPS uses the native spin hyperedges. For CATN we set its boundary-MPS cap and merged-index cap to the same value, \(D_{\max}=\chi\). HAC uses its late-compression implementation with a tree-gauge radius of two and a contraction path optimized for each \(\chi\).

\subsection{Contraction accuracy and scalability}
Figure~\ref{fig:ising} shows the results for the \(6\times6\times6\) Ising model. We quantify the relative error by
\begin{equation}
 \epsilon_Z=\frac{|\widehat Z-Z_{\mathrm{ref}}|}{|Z_{\mathrm{ref}}|}.
 \label{eq:partition-error}
\end{equation}

These results demonstrate that CT-bMPS can accurately approximate the partition function. On the SA-optimized tree, increasing $\chi$ systematically improves the accuracy of both ordinary and environment CT-bMPS. Environment updates give smaller errors at the same $\chi$, at the cost of additional contraction time and memory. The reduction in relative error reaches approximately three orders of magnitude compared with ordinary CT-bMPS on the same tree.

The contraction tree has a substantial effect on this accuracy--cost tradeoff. At the same bond dimension, replacing the $xyz$-sweep tree with the SA-optimized tree can reduce contraction time by roughly an order of magnitude and relative errors by several orders of magnitude. The peak process memory can also be substantially reduced for environment CT-bMPS. The tree determines which intermediate tensors must be represented by boundary MPS. These results show that a geometrically intuitive contraction order does not necessarily give the best approximation, even on a regular lattice. A tree obtained by optimizing exact-contraction FLOPs can yield much more accurate scalar estimates at the same $\chi$ and substantially reduce the contraction time. The ability to use general contraction trees is therefore a practical advantage of CT-bMPS.

\begin{figure}[t]
\centering
\includegraphics[width=\linewidth]{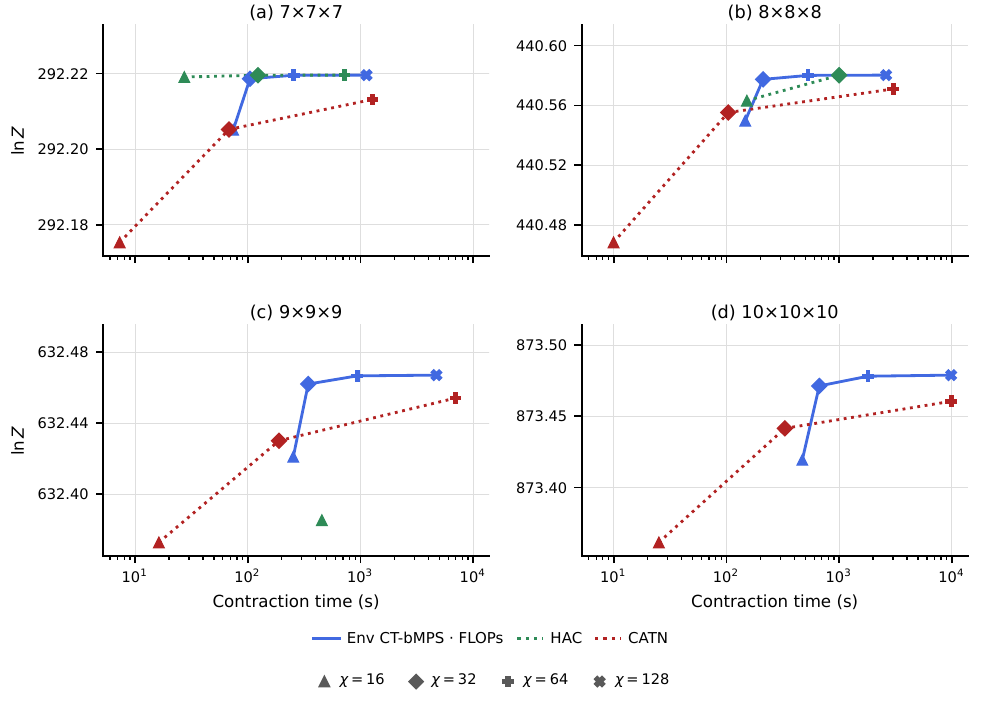}
\caption{Computed \(\ln Z\) against numerical contraction time for cubic Ising lattices of side \(L=7,8,9,10\) with open boundary conditions. Independent exact references are unavailable at these sizes. HAC has no completed point at \(L=10\).}
\label{fig:ising-large}
\end{figure}

\begin{figure}[t]
\centering
\includegraphics[width=\linewidth]{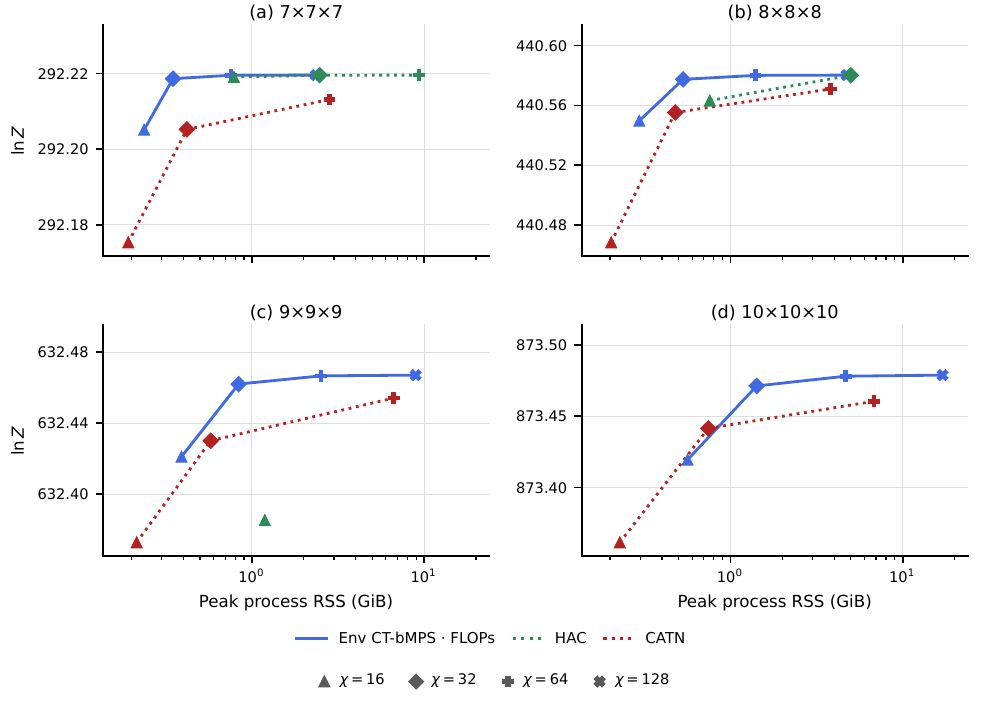}
\caption{Computed $\ln Z$ against measured peak process RSS for the same $L=7,8,9,10$ Ising benchmarks as Fig.~\ref{fig:ising-large}.}
\label{fig:ising-large-memory}
\end{figure}

At $L=6$, environment CT-bMPS achieves high accuracy comparable to HAC with similar contraction time and memory. Its advantage becomes clearer when the comparison is extended to larger lattices. Figures~\ref{fig:ising-large} and~\ref{fig:ising-large-memory} show the computed $\ln Z$ as a function of contraction time and peak memory usage, respectively, for $L=7,8,9,10$. Since independent exact references are unavailable, we examine the convergence of the computed $\ln Z$ with increasing bond dimension. At $L=7$, HAC approaches the stable value faster than environment CT-bMPS, while at $L=8$, environment CT-bMPS reaches comparable values using less contraction time and memory. For $L=9$ and $L=10$, memory limits prevent HAC from being run at large bond dimensions, whereas CT-bMPS continues to produce stable approximations as the bond dimension increases. These results demonstrate the scalability of CT-bMPS in both contraction time and memory usage. CATN also remains executable at all four sizes, but requires more contraction time and memory than CT-bMPS to approach the converged values.

\subsection{Random tensor networks with signed entries}
Finally, to investigate which TNs can be accurately approximated by CT-bMPS, we benchmark random TNs with different tensor-entry distributions. We contract random TNs on a \(5\times5\times5\) cubic graph with bond dimension \(D=2\). Each tensor entry is independently drawn from
\begin{equation}
 T^{(v)}_{\boldsymbol i}(\lambda)
 \sim\operatorname{Uniform}[\lambda,1].
 \label{eq:random-tensor-family}
\end{equation}
We define the contraction tree and global index order in the same way as for the Ising benchmarks. We run ordinary and environment CT-bMPS at \(\chi=32\), using five environment sweeps.

\begin{figure}[t]
\centering
\includegraphics[width=0.50\linewidth]{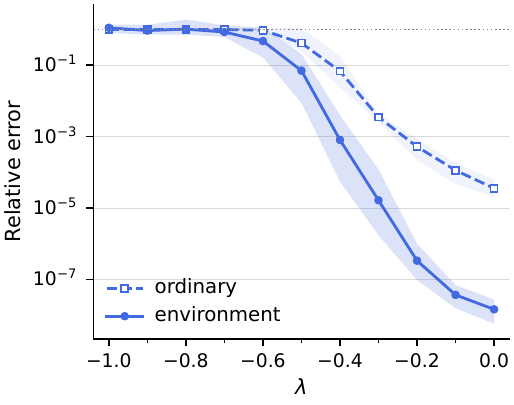}
\caption{Dependence on tensor-entry distribution for closed random tensor networks on an open-boundary cubic lattice with \(L=5\), \(D=2\), and \(\chi=32\). Lines are median relative scalar errors across twenty paired instances per \(\lambda\); shaded bands show the 25th--75th percentiles. Ordinary and environment calculations use the same fixed contraction tree and global order.}
\label{fig:random-tn-lambda}
\end{figure}

Figure~\ref{fig:random-tn-lambda} shows the performance of CT-bMPS for different values of \(\lambda\). The error begins to decrease markedly around \(\lambda\simeq-0.7\), showing a crossover between difficult and approximable TNs, which was also observed in previous work~\cite{gray2024}. As \(\lambda\) approaches \(-1\), the proportions of positive and negative entries become nearly equal, and strong cancellations make accurate approximation difficult: both methods retain relative errors of order one. For larger \(\lambda\), CT-bMPS achieves small errors, and environment updates substantially improve the accuracy at the same bond dimension.

\begin{figure}[t]
\centering
\includegraphics[width=\linewidth]{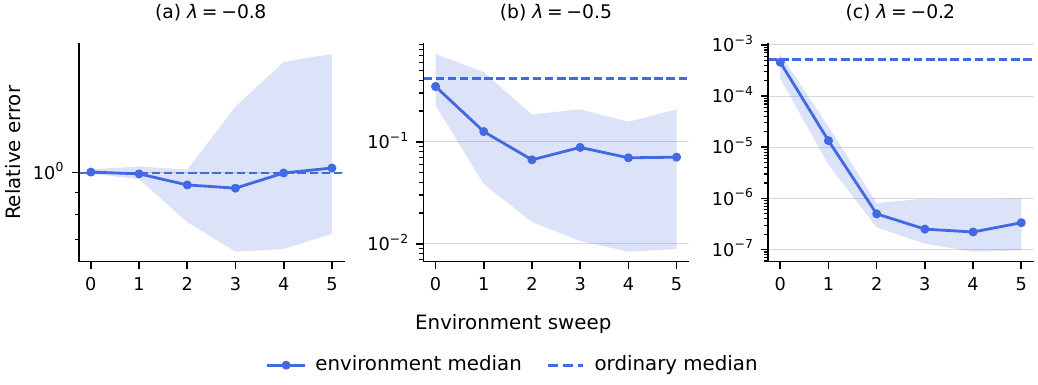}
\caption{Sweep dependence for the same twenty-instance random-network ensembles at \(\lambda=-0.8,-0.5,-0.2\). Solid lines and bands give medians and interquartile ranges of the environment error. Dashed horizontal lines give the corresponding ordinary medians.}
\label{fig:random-tn-sweeps}
\end{figure}
\FloatBarrier

Figure~\ref{fig:random-tn-sweeps} shows how the error changes with the number of environment sweeps. At \(\lambda=-0.8\), additional sweeps leave the median relative error of order one. At \(\lambda=-0.5\) and \(-0.2\), the sweeps substantially reduce the error. In the approximable regime, CT-bMPS works well, and environment updates further improve its accuracy without increasing \(\chi\).

\section{Quantum circuit simulation}
\label{sec:benchmark-ibm}

\subsection{Circuit model and classical simulation methods}

We next apply CT-bMPS to quantum circuit simulation by evaluating observables as closed TN contractions. We consider the kicked Ising circuits implemented on IBM's Eagle processor with 127 qubits~\cite{kim2023}, a benchmark studied extensively by classical simulation methods. Starting from \(\lvert 0\rangle^{\otimes 127}\), we apply a transverse rotation and then commuting Ising interactions on the 144 edges of the heavy-hex graph \(G\) in each time step:
\begin{equation}
 U(\theta_h)=
 \left[\prod_{(i,j)\in E(G)}\exp\!\left(\frac{i\pi}{4}Z_iZ_j\right)\right]
 \left[\prod_{j\in V(G)}\exp\!\left(-\frac{i\theta_h}{2}X_j\right)\right].
 \label{eq:ibm-unitary}
\end{equation}
The desired scalar is the expectation value of an observable after repeated applications of \(U(\theta_h)\).
We study the angle dependence for \(0\leq\theta_h\leq\pi/2\). The five-step example measures the weight-17 Pauli string \(P_{17}\) studied in Ref.~\cite{kim2023}, whose light cone contains 68 qubits. The deeper example measures \(Z_{62}\) after twenty complete steps and involves all 127 qubits.

Numerous classical simulation methods have been developed for this circuit~\cite{patra2024,liao2023,anand2023,begusic2023spd,rudolph2023lowesa,kechedzhi2024}. The original study used MPS and isometric tensor network states (isoTNS) for classical simulations~\cite{kim2023}. Tindall \emph{et al.} represent the quantum state as a PEPS on the heavy-hex lattice and use belief propagation to gauge the TN during time evolution, obtaining accurate estimates of the observables~\cite{tindall2024}. Begu\v{s}i\'{c} \emph{et al.} developed a mixed-picture (MIX) method that evolves both the quantum state and the observable as TNs and uses belief propagation to evaluate the resulting expectation value~\cite{begusic2024}. We use their MIX results as an approximate classical reference for the twenty-step circuit.

\subsection{Tensor network construction and contraction tree selection}

We first remove gates outside the observable's light cone. We decompose each two-qubit gate into a pair of rank-three tensors connected by a bond of dimension two. We absorb the single-qubit rotations, initial states, and observable into these tensors. We then fold the TN by combining corresponding tensors from the ket and bra circuits and grouping each pair of matching indices into a single index~\cite{banuls2009,hastings2015}.

We apply environment CT-bMPS to the folded TNs, defining the global index order geometrically. For nonuniform TNs such as this circuit, neither a geometric sweep nor a tree optimized for exact-contraction FLOPs necessarily gives the most accurate approximation. For the twenty-step circuit, we therefore generate multiple candidate trees through independently seeded SA searches for low exact-contraction FLOPs. We screen these candidates at a small bond dimension for their ability to retain the observable signal and use a promising tree for the main calculations at \(\chi=64\). Developing contraction tree optimization for such nonuniform TNs remains future work.

\subsection{Simulation results}

\begin{figure}[t]
\centering
\includegraphics[width=\linewidth]{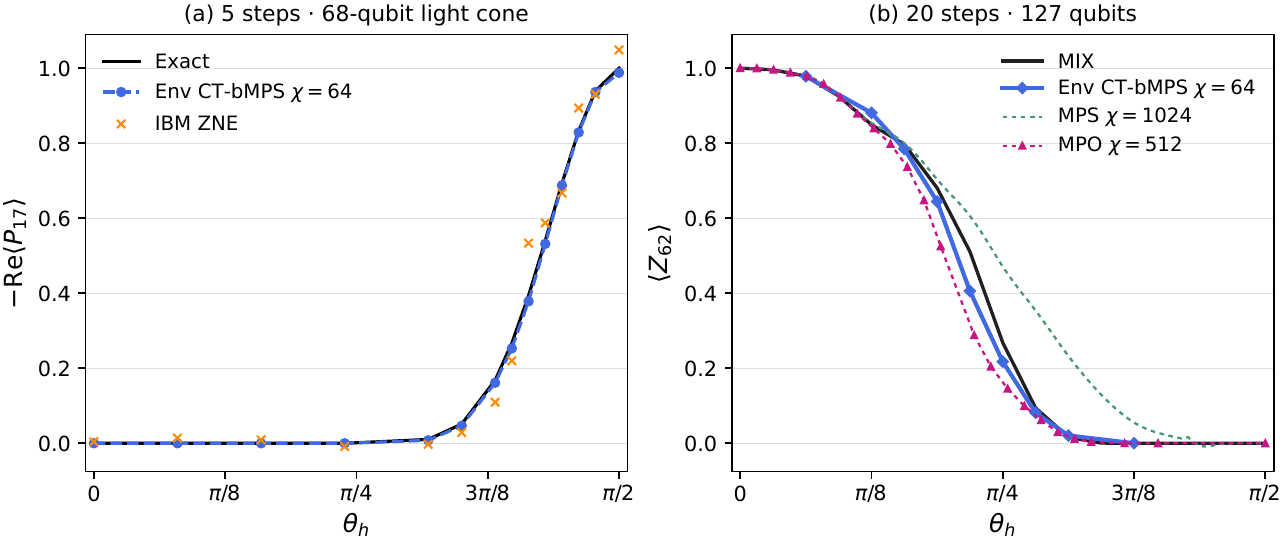}
\caption{Kicked Ising dynamics on the IBM heavy-hex circuit. (a) The weight-17 observable after five steps, restricted to its 68-qubit light cone. CT-bMPS uses \(\chi=64\) and one environment sweep; the black curve is an exact-contraction reference, and orange crosses show IBM zero-noise-extrapolated measurements~\cite{kim2023}. (b) \(\langle Z_{62}\rangle\) after twenty steps on 127 qubits. Environment CT-bMPS uses \(\chi=64\) and ten sweeps. The approximate references are MIX~\cite{begusic2024}, MPS with \(\chi=1024\)~\cite{kim2023}, and MPO with \(\chi=512\)~\cite{anand2023}.}
\label{fig:ibm}
\end{figure}

Figure~\ref{fig:ibm} shows the results of quantum circuit simulation using CT-bMPS. For the smaller five-step circuit in panel (a), CT-bMPS agrees well with the exact results. For the larger twenty-step circuit in panel (b), environment CT-bMPS does not fully reproduce the MIX reference, but captures its overall angle dependence and sharp decrease in the observable. These results demonstrate the applicability of CT-bMPS to quantum circuit simulation.

In particular, environment CT-bMPS with only \(\chi=64\) agrees more closely overall with the MIX reference than the MPS simulation with \(\chi=1024\)~\cite{kim2023} and the MPO simulation with \(\chi=512\)~\cite{anand2023}. Since the details of these algorithms differ, their bond dimensions alone do not provide a direct comparison of total computational cost. Nevertheless, such small boundary bond dimensions can substantially reduce the arithmetic cost of local updates. For the folded TN with physical dimension four, ordinary truncation uses matrices of size at most \(\chi^2\times4\chi\). Moreover, the environment-assisted SVD acts on smaller matrices of size at most \(\chi\times4\chi\). These local contractions and SVDs are natural candidates for GPU acceleration, which could reduce runtime and make larger bond dimensions accessible, potentially improving the approximation accuracy.

\section{Quantum error correction decoding}
\label{sec:benchmark-decoding}

\subsection{Logical probabilities as closed hyperedge contractions}

Our final application is the decoding of quantum low-density parity-check (qLDPC) codes. We express logical-class probabilities as closed TN contractions and apply CT-bMPS to surface and color codes under circuit-level noise. We describe independent binary faults $x_q\in\{0,1\}$ by a detector error model (DEM)~\cite{gidney2021stim,piveteau2024}, with probabilities $\mu_q(0)=1-p_q$ and $\mu_q(1)=p_q$. For a single logical bit, the unnormalized probability of class $\ell$ is
\begin{equation}
 Z_\ell(s)=\sum_{x:Hx=s,\,Lx=\ell}\prod_q\mu_q(x_q).
 \label{eq:dem-sector}
\end{equation}
Here $s$ is the observed detector syndrome, $H$ maps faults to detector outcomes, and $L$ maps faults to the logical bit, with arithmetic modulo two. Logical maximum-likelihood decoding (MLD) selects the class with the larger probability. This differs from most-likely-error (MLE) decoding, which selects one compatible fault configuration,
\begin{equation}
 x_{\mathrm{MLE}}\in\mathop{\mathrm{argmin}}_{x:Hx=s}
 \sum_q x_q\log\frac{1-p_q}{p_q},
 \label{eq:dem-mle}
\end{equation}
and returns its logical class. Degeneracy matters in Eq.~\eqref{eq:dem-sector}: many individually unlikely configurations can together outweigh the class containing the most likely individual configuration. Summing logical-class probabilities can therefore improve decoding over selecting an individual fault configuration.

We compute the unnormalized Fourier coefficients of the logical-sector weights,
\begin{equation}
 \begin{aligned}
 F_0(s)&=Z_0(s)+Z_1(s),\\
 F_1(s)&=Z_0(s)-Z_1(s)
 =\sum_{x:Hx=s}(-1)^{Lx}\prod_q\mu_q(x_q).
 \end{aligned}
 \label{eq:dem-f1}
\end{equation}
The decision is $\widehat\ell=0$ for $F_1>0$ and $\widehat\ell=1$ for $F_1<0$. The quantity $F_1$ can be expressed as the contraction of a single TN, as shown below. The parity constraint at each detector $a$ is represented by the tensor
\begin{equation}
 A_a(\{x_q:q\in N(a)\})=
 \mathbf{1}\!\left[\bigoplus_{q\in N(a)}x_q=s_a\right],
 \label{eq:dem-parity}
\end{equation}
where $N(a)$ is the set of faults that flip detector $a$, and $s_a$ is its observed binary outcome. Each fault is a binary index shared by all incident detector tensors and carries the factor $\mu_q(x_q)(-1)^{L_qx_q}$. Contracting these factors with the detector parity tensors gives $F_1(s)$ as a closed TN with hyperedges.

DEM TNs contain many hyperedges, and each detector tensor can have a large number of indices. A dense representation of a detector tensor with $m$ binary indices requires $2^m$ entries, making a direct application of methods that retain dense tensors, such as HAC~\cite{gray2024}, to this representation prohibitively expensive. In contrast, each parity tensor $A_a$ has an exact MPS representation with bond dimension two. CT-bMPS therefore takes these parity MPS directly as inputs and represents intermediate tensors as MPS throughout the contraction. At fixed MPS bond dimension, their storage grows linearly with the number of indices. For compression, each boundary fault index carries a square-root prior weight $\sqrt{\mu_q(x_q)}$. A factor $\mu_q(x_q)^{-1/2}$ removes the duplicated weight when a shared fault index remains on the outgoing boundary.

The global index order follows the physical fault locations. For each fault, we collect the distinct spatial-and-time centers of all corresponding noise-gate locations. We define the global index order by sorting the faults according to the mean coordinates of these centers. In the environment CT-bMPS calculations below, we read $F_1$ at a fixed root during the final sweep.

\subsection{Numerical accuracy and computational cost}

We first examine the numerical accuracy of $F_1$ itself for the distance-five surface code, for which exact contraction references are available. We use eight fixed shots at $p=0.009$. The DEM contains 1,679 fault mechanisms, and its TN representation consists of 120 detector-parity tensors connected by 1,607 binary fault indices after single-detector faults are absorbed locally. Fault indices can be shared by up to four tensors, giving an irregular hypergraph rather than a regular lattice TN. We measure the error as
\begin{equation}
 \epsilon_{F_1}(s)=
 \frac{|\widehat F_1(s)-F_1(s)|}{F_0(s)}.
 \label{eq:dem-f1-error}
\end{equation}

\begin{figure}[t]
\centering
\includegraphics[width=\linewidth]{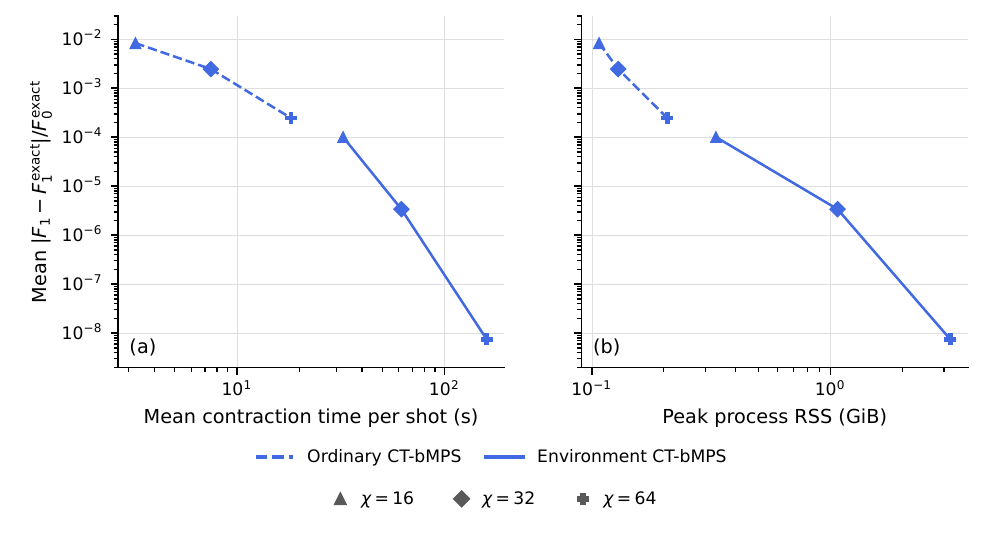}
\caption{Numerical contraction accuracy and computational cost for surface code shots at $d=5$, $p=0.009$. The ordinate is the mean of Eq.~\eqref{eq:dem-f1-error}; the horizontal axes show mean contraction time per shot and maximum process peak RSS over the shots. Markers indicate $\chi$; dashed and solid lines distinguish ordinary and environment CT-bMPS. Environment calculations use $\chi_0=\chi$ and five sweeps for all bond dimensions.}
\label{fig:dem-f1-cost}
\end{figure}

Figure~\ref{fig:dem-f1-cost} compares numerical accuracy with contraction time and memory usage. CT-bMPS accurately approximates $F_1$, with the plotted error decreasing as the bond dimension increases. Environment updates yield substantially higher accuracy at the same bond dimension, at the cost of additional time and memory. These observations demonstrate that CT-bMPS can approximate the DEM contractions required for decoding.

Contraction accuracy and decoding performance are not in one-to-one correspondence, because the decoding decision depends only on the sign of $F_1$. Improving its numerical value does not change the decision as long as the sign is preserved. Conversely, when the absolute approximation error exceeds $|F_1|$, it can reverse the sign, changing the inferred logical class and potentially causing a decoding error.

\subsection{Surface and color code decoding}
\label{sec:dem-distance-context}

We evaluate the logical error rate of the rotated surface code at $p=0.009$ for distances $d=3,5,7$, with $d$ syndrome rounds. We use ordinary CT-bMPS with $\chi=32$ at $d=3,5$ and environment CT-bMPS with $\chi=16$ and two sweeps at $d=7$. The $d=7$ TN forms a large irregular hypergraph with 336 detector-parity tensors and 5,833 binary fault indices, including 3,931 hyperedges.

We compare with the published TN decoder of Piveteau \emph{et al.}~\cite{piveteau2024} and the MLE decoder Tesseract~\cite{beni2025}. The former embeds the detector TN into a cubic lattice and contracts it by a PEPS sweep with simple-update compression, followed by a boundary MPS contraction of the remaining two-dimensional TN. The PEPS intermediate does not generally admit the canonical form available for an MPS, which can limit the accuracy of local truncations. CT-bMPS instead keeps intermediate tensors in canonical MPS representations. Tesseract searches for a most-likely compatible fault configuration using A* search with pruning. For both codes, we use \texttt{det\_beam}$=15$ and \texttt{pqlimit}$=2\times10^5$, with beam climbing enabled and 16 detector orders. Our TN and Tesseract results use the same shots within each code and distance.

\begin{figure}[t]
\centering
\includegraphics[width=0.66\linewidth]{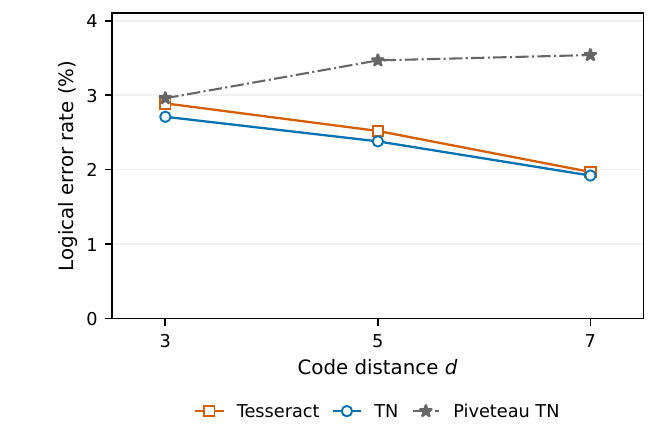}
\caption{Rotated surface code logical error rates at $p=0.009$ with $d$ syndrome rounds. Our TN and Tesseract~\cite{beni2025} results use the same shots at each distance. TN uses ordinary CT-bMPS with $\chi=32$ at $d=3,5$ and environment CT-bMPS with $\chi=16$ and two sweeps at $d=7$. The Piveteau TN curve is digitized from Fig.~1(d) of Ref.~\cite{piveteau2024} and uses independent shots.}
\label{fig:codes}
\end{figure}

Figure~\ref{fig:codes} shows the surface code results. Our TN decoder achieves lower logical error rates than the published TN decoder, with a substantial improvement at $d=5$ and $d=7$, demonstrating the effectiveness of our approximation method. The observed logical error rates are also lower than Tesseract's, but the differences are small. These results support the use of logical-class probability sums for decoding, while showing that the gain over the tested MLE decoder is modest.

We also consider the superdense color code memory-$X$ circuit~\cite{gidneyjones2023} at $p=0.0025$. TN decoding is exact at $d=3,5$, while $d=7$ uses environment CT-bMPS with $\chi=16$ and five sweeps. The $d=7$ TN contains 252 detector-parity tensors and 9,869 binary fault indices, including 9,015 hyperedges. Although it contains fewer tensors than the surface code TN, it has more fault indices and higher-order hyperedges, which connect up to seven tensors rather than four. Its connectivity is therefore more complex.

\begin{figure}[t]
\centering
\includegraphics[width=0.66\linewidth]{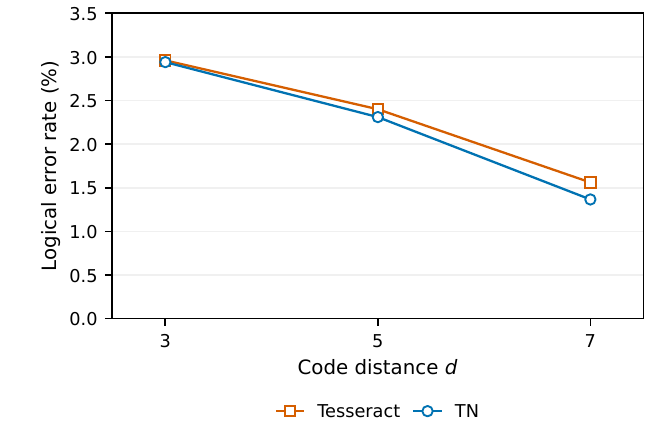}
\caption{Superdense color code memory-$X$ logical error rates at $p=0.0025$ and $d$ rounds. TN and Tesseract use the same shots at each distance. TN is exact at $d=3,5$ and uses uniform $\chi=16$ with five full-environment sweeps at $d=7$.}
\label{fig:dem-color}
\end{figure}

Figure~\ref{fig:dem-color} shows the color code results. As for the surface code, TN decoding gives slightly lower observed logical error rates than Tesseract at each displayed distance. The same CT-bMPS contraction procedure thus applies to a code with different geometry and more complex fault connectivity.

The TN calculations required substantially longer runtimes than Tesseract. Nevertheless, CT-bMPS provides an approximate MLD decoder that can be used to evaluate logical-class probabilities and investigate QEC performance. Information specific to decoding enters through the TN representation and its probability weights, while the contraction and truncation procedure remains general. Once a DEM is expressed in this form, the same framework can be applied across different codes and geometries without designing a separate approximation scheme for each code and noise model.

\section{Discussion}
\label{sec:discussion}

In this work, we have proposed CT-bMPS, a general approximate contraction method for closed TNs. The method follows a contraction tree and represents intermediate boundary tensors as MPS. Using a backward MPS from the opposite side of each tree edge as an environment can further improve accuracy at the same bond dimension. Ordinary and environment CT-bMPS both have time and memory costs that grow polynomially with the TN size and the MPS bond dimension. Our benchmarks demonstrate accurate and scalable approximation across several models, with improvements in accuracy and accessible system sizes compared with existing approximate contraction methods.

One advantage of CT-bMPS is its generality and scalability. Once a problem is expressed as a closed TN and a contraction tree and global index order are specified, the same boundary MPS contraction and truncation procedures can be applied. As in exact TN contraction, a common framework can therefore be reused without redesigning the approximate scheme for each problem. Whereas HAC represents intermediate tensors densely, CT-bMPS keeps them in MPS form. At fixed MPS bond dimension, the cost grows polynomially rather than exponentially with the number of boundary indices, allowing execution on larger TNs. Beyond the present benchmarks, this framework may be useful for finite TNs arising in statistical mechanics, quantum many body physics, quantum circuit simulation, and combinatorial optimization.

In our benchmarks, environment updates substantially improve approximation accuracy at the same MPS bond dimension. Achieving a target accuracy with smaller bond dimensions can reduce both the arithmetic cost and the workspace required for local tensor contractions and SVDs. However, the environment procedure also requires storing MPS messages in both directions on every tree edge, and this message bank can account for a substantial memory overhead. Our current implementation uses a simple sequence of local contractions and SVDs. Combining CT-bMPS with more sophisticated compression techniques, such as those developed for products of MPS and MPO~\cite{schollwoeck2011,camano2026}, may improve the local approximation and reduce the storage required for the MPS message bank. The small matrices involved in the local operations also motivate GPU implementations. Developing these compression strategies and efficient GPU implementations while preserving the polynomial cost guarantee remains future work.

The accuracy of CT-bMPS depends strongly on the contraction tree and global index order. We have used SA to optimize exact-contraction FLOPs and geometric coordinates to define the index order. These choices provide a practical starting point, but do not directly optimize approximation accuracy. For the kicked Ising simulation, we screened several candidate trees at a small bond dimension before selecting one for the main calculation. An important next step is to develop methods that select contraction trees and index orders according to how well the resulting intermediate tensors can be approximated. In addition to exact-contraction FLOPs and intermediate tensor sizes, singular value spectra and entanglement entropies of intermediate MPS may provide useful information for this purpose~\cite{hikihara2023}. Incorporating this information into the optimization of contraction trees and global index orders may further improve the accuracy and computational efficiency of CT-bMPS.

\section*{Acknowledgements}

We thank Itai Arad for useful discussions. This work is supported by the Ministry of Education Singapore under grant No. SKI 2021\_07\_03, National Quantum Computing Hub translational fund No. W24Q3D0002, MTC Young Individual Research Grant No. M25N8c0120 and Advanced Quantum Algorithms and Solutions (AQAS) Grant No. S25Q9DA001/02.

\FloatBarrier
\setlength{\bibitemsep}{3pt plus 1pt minus 1pt}
\printbibliography

\end{document}